\documentstyle[11pt,aaspp,flushrt]{article}

\catcode`\@=11 
\def\@versim#1#2{\vcenter{\offinterlineskip
        \ialign{$\m@th#1\hfil##\hfil$\crcr#2\crcr\sim\crcr } }}
\newcommand{\beq}{\begin{equation}}
\newcommand{\eeq}{\end{equation}}
\def\lsim{\mathrel{\mathpalette\@versim<}}
\def\gsim{\mathrel{\mathpalette\@versim>}}
\def\vp{v_\phi}
\def\la{\langle}
\def\ra{\rangle}

\begin{document}
\title{Convection-Dominated Accretion Flows}
\author{Eliot Quataert\footnote{Chandra Fellow} and Andrei Gruzinov}
\affil{Institute for Advanced Study, School of Natural Sciences, Olden Lane, 
Princeton, NJ 08540; eliot@ias.edu; andrei@ias.edu}

\medskip

\begin{abstract}

Non-radiating, advection-dominated, accretion flows are convectively
unstable in the radial direction.  We calculate the two-dimensional
($r-\theta$) structure of such flows assuming that (1) convection
transports angular momentum inwards, {\it opposite} to normal
viscosity and (2) viscous transport by other mechanisms (e.g.,
magnetic fields) is weak ($\alpha \ll 1$).  Under such conditions
convection dominates the dynamics of the accretion flow and leads to a
steady state structure that is marginally stable to convection.  We
show that the marginally stable flow has a constant temperature and
rotational velocity on spherical shells, a net flux of energy from
small to large radii, {\it zero} net accretion rate, and a radial
density profile of $\rho \propto r^{-1/2}$, flatter than the $\rho
\propto r^{-3/2}$ profile characteristic of spherical accretion flows.
This solution accurately describes the full two-dimensional structure
of recent axisymmetric numerical simulations of advection-dominated
accretion flows.

\end{abstract}

\section{Introduction}

Analytical calculations of the structure of non-radiating,
advection-dominated, accretion flows (ADAFs) have shown that they are
convectively unstable in the radial direction (e.g., Begelman \& Meier
1982; Narayan \& Yi 1994).\footnote{Note that this is very different
from thin accretion disks which can be vertically convective.}  In the
absence of radiation, the entropy of the gas increases as it accretes
(due to viscous dissipation) and the sub-Keplerian rotation of the
flow is insufficient to stabilize this unstable entropy gradient.
This conclusion has been confirmed by several numerical simulations
(Igumenshchev, Chen, \& Abramowicz 1996; Igumenshchev \& Abramowicz
1999, hereafter IA; Stone, Pringle, \& Begelman 2000, hereafter SPB).

Narayan \& Yi (1994; 1995) argued that convection was unlikely to
significantly modify the structure of ADAFs, essentially because the
inflow time of the gas would be shorter than the characteristic
convective turnover time (i.e., advection would overwhelm convection).
Their analysis assumed that convection behaved like ``normal''
viscosity, transporting angular momentum and energy from small to
large radii in the flow.


A number of treatments of convection in thin accretion disks, however,
have argued that convection does not act like normal viscosity;
instead it transports angular momentum inwards, opposite to what is
needed to allow accretion (Ryu \& Goodman 1992; Stone \& Balbus 1996;
Balbus, Hawley, \& Stone 1996).  Narayan, Igumenshchev, \& Abramowicz,
(2000; hereafter NIA) have shown that, if present in ADAFs, this
inward angular momentum transport by convection can dramatically
modify the structure of the accretion flow, essentially suppressing
accretion.  In this paper we reach the same conclusion as NIA via a
different analysis; we also find the full two-dimensional structure of
the flow (NIA considered height integrated flows).

We assume that convection transports angular momentum inwards and that
outward transport by other mechanisms (e.g., magnetic fields) is weak
in the sense that the induced radial velocities are highly subsonic
(the Shakura-Sunyaev dimensionless viscosity or its magnetic analogue
is small; $\alpha \ll 1$). In this case the steady state structure of
the accretion flow must, statistically speaking, approach marginal
stability to convection.  This is because, in the absence of marginal
stability, the characteristic convective turnover time is of order the
dynamical time, much shorter than the inflow time of the gas (because
$\alpha \ll 1$).  Convection therefore dominates the flow structure
and forces it to marginal stability (just like in the solar convection
zone).

In the next section (\S2) we calculate the two-dimensional
($r-\theta$) structure of an axisymmetric, radially self-similar
accretion flow that is marginally stable to convection.  This provides
us with the angular structure of the flow.  In \S3 we argue that the
only marginally stable accretion flow which satisfies mass, angular
momentum, and energy conservation is one for which there is no
accretion ($\dot M = 0$), but a net flux of energy from small to large
radii.\footnote{In the self-similar limit, a finite energy flux is
generated by a ``black hole'' of infinitesimal radius which accretes
an infinitesimal amount of matter.}  This is in contrast to standard
ADAF solutions in which there is no net energy flux, but a finite
accretion rate (see, e.g., Narayan \& Yi 1994; Blandford \& Begelman
1999).  In \S4 we compare our solution to recent numerical simulations
and discuss its relevance to ``real'' ADAFs.

\section{Marginal Stability:  the Angular Structure}

We consider stability to local, axisymmetric, adiabatic,
perturbations.  Entropy gradients in ADAFs tend to destabilize
perturbations while rotation stabilizes them.  The net stability
criterion can be found by requiring that the projection of the
buoyancy and centrifugal forces on the displacement vector of a mode
is negative (so there is a net restoring force for any perturbation);
this yields the ``Hoiland'' criterion (e.g., Tassoul 1978; Begelman \&
Meier 1982)
\begin{equation}
({\bf \nabla} s \cdot {\bf dr}) ({\bf g} \cdot {\bf dr}) - {2 \gamma
\vp \over R^2} \left({\bf \nabla } [\vp R] \cdot {\bf dr}\right) dR\leq 0. 
\label{stability}
\end{equation}
In equation (\ref{stability}), $R=r\sin \theta$ is the cylindrical
radius, ${\bf dr} = dr \hat r + r d \theta \hat \theta$ is the
displacement vector, $s = \ln(p) - \gamma \ln(\rho)$ is $(\gamma - 1)$
times the entropy, ${\bf g} = - \hat r v_K^2/r + \hat R \vp^2/R$ is
the effective gravity, and $v_K = r^{-1/2}$ and $\vp$ are the
Keplerian and rotational velocities, respectively (taking $GM = 1$,
where $M$ is the mass of the central object).

Away from the boundaries, there is only one characteristic length
scale in an ADAF, the distance from the central object ($r$).  We
therefore consider radially self-similar flows in which the dynamical
variables are power laws in radius times functions of $\theta$, i.e.,
$\rho =r^{-n} \rho (\theta )$, $c^2 = r^{-1} c^2(\theta)$, $p = r^{-1
- n}p(\theta )$, and $\vp^2=r^{-1} \vp^2 (\theta )$; $c^2$ is the
isothermal sound speed (i.e., temperature) and $p = \rho c^2$ is the
pressure.  The radial scalings of the temperature and rotational
velocity are set by the $1/r$ gravitational potential.  For reasons
that will become clear in \S3 we allow the density to be an arbitrary
power law in radius, $\rho \propto r^{-n}$, rather than requiring the
usual spherical flow value of $n = 3/2$.

In the Appendix we show that, for a given $n$ and $\gamma$, the
requirement that the flow be marginally stable to convection for all
$\theta$ determines its structure.\footnote{Marginal stability means
that the quadratic form given by equation (\ref{stability}) is
negative along all directions but one, along which it is equal to
zero; see the discussion below equation (\ref{marg}) in the Appendix.}
The temperature and rotational velocity are constant on spherical
shells, i.e., independent of $\theta$, with values

\begin{equation}
c^2={v^2_K\over {\gamma +1\over \gamma -1}-n} \ \ \  {\rm and}  ~~~~\vp^2=2 
v^2_K 
\left({{1\over \gamma -1} -n\over
{\gamma +1\over \gamma -1}-n}\right). \label{solution}
\end{equation}

The density is then a power law in $\sin\theta$, with \beq \rho(\theta)
\propto (\sin \theta )^{2({1\over \gamma -1}-n)}. \eeq

Note that in the above solution the surfaces of constant entropy and
angular momentum coincide, and are given by $r \sin^2 \theta = {\rm
constant}$.  Such ``gyrentropic'' solutions are expected for marginal
stability to convection, and have been discussed previously in the
literature (e.g., Paczynski \& Abramowicz 1982; Blandford,
Jaroszynski, \& Kumar 1985).

\section{Implications of Marginal Stability:  the Radial Structure}

The marginally stable solution derived in the previous section
describes the angular structure of the flow for a given $\gamma$ and
$n$.  Here we argue that the requirement of marginal stability,
together with mass, angular momentum, and energy conservation, also
uniquely fixes the radial density power law, $n$.  We first give a
physical argument and then elaborate on it with some non-rigorous
mathematics.

Since radiation is assumed negligible, we write the conservation laws
as \beq {d \bar F \over d r } = 0, \ \ {\rm where} \ \ \bar F \equiv
\int^\pi_0 d \theta \sin \theta F. \eeq $F$ represents the
time-averaged radial flux of energy ($F_E$), angular momentum ($F_L$),
and mass ($F_M$) at a given $r$ and $\theta$, while $\bar F$ is the
corresponding angle-integrated flux (i.e., the net flux through a
spherical shell of radius $r$).

In standard ADAF models, $n = 3/2$; as discussed below, this
corresponds to a finite flux of mass onto the central object ($\bar
F_M < 0$), but zero net angular momentum and energy flux, i.e., $\bar
F_E = \bar F_L = 0$ (Narayan \& Yi 1994; Blandford \& Begelman 1999).
That such a solution is possible can be understood as follows.
Viscosity gives rise to a net torque on the gas which drives it
inwards ($\bar F_M < 0$).  The outward angular momentum flux due to
this torque is exactly balanced by the inward flux of angular momentum
due to the bulk motion of the gas; this is how $\bar F_L = 0$.
Associated with the outward flux of angular momentum due to the
viscous stress is an outward flux of energy due to the work done by
the stress on the gas.  The net energy flux is zero, i.e., $\bar F_E =
0$, because the outward energy flux due to viscosity is exactly
balanced by the inward flux of energy due to the bulk motion of the
gas.  The latter is given by $F_M Be$, where $Be$ is the Bernoulli
constant, \beq Be = {1 \over 2} v^2 + {\gamma \over \gamma - 1} c^2 -
v_K^2, \eeq where $v^2 = \vp^2 + v_r^2 + v_\theta^2$.  For standard
ADAF solutions, $Be$ must be positive and relatively large (i.e., a
reasonable fraction of $v^2_K$) in order for the inward mechanical
flux of energy to balance the outward viscous flux (Narayan \& Yi
1994; Blandford \& Begelman 1999).

From equation (\ref{solution}) one can readily show that the Bernoulli
constant satisfies $ Be \approx 0$ {\it for all $\theta, \gamma$ and
$n$} in the marginally stable state.\footnote{An alternative proof
that $Be \approx 0$ is as follows: the marginally stable flow is
gyrentropic, with surfaces of constant angular momentum and entropy
coinciding.  Under these conditions, one can show that the surfaces of
constant $Be$ coincide with the surfaces of constant angular momentum
and entropy.  Since $\vp$ and $c$ are independent of $\theta$, $Be$ is
constant on spherical shells.  Angular momentum and entropy are,
however, constant on surfaces defined by $r \sin^2 \theta = {\rm
constant}$.  Thus $Be$ must be zero at marginal stability.}  It is
therefore not possible to have a normal ADAF, i.e., an $n = 3/2$
solution.  There is no inward mechanical flux of energy to balance the
outward flux due to viscosity.

Physically, the above argument shows that a marginally stable flow
must have a net flux of energy to large radii. This uniquely fixes $n$
for a given viscosity prescription since there is only one $n$ for
which the outward flux of energy is independent of radius, as is
required in steady state.  Modeling viscosity via an $r-\phi$
component of the stress tensor, the viscous energy flux is $\propto
r^2 \nu \rho \vp^2/r$, where $\nu$ is the kinematic viscosity.  For a
self-similar flow, the viscous flux is $\propto \nu \rho \propto \nu
r^{-n}$ and so if $\nu \propto r^\beta$, a finite energy flux requires
$n = \beta$.  For a Shakura-Sunyaev viscosity prescription, $\nu =
\alpha c r \propto r^{1/2}$, so that $n = 1/2$.  In this marginally
stable solution with a finite energy flux, $\bar F_M = 0$, i.e., there
is no accretion; matter simply circulates in convective eddies.  This
is possible because the inward transport of angular momentum by
convection balances the outward transport by viscosity.

To elaborate on this discussion, we construct specific expressions for
$F_L$ and $F_E$.  This elaboration is intended solely to motivate the
various signs utilized in the above argument.  We neither can nor do
construct a fully satisfactory model for the energy and angular
momentum fluxes in a turbulent/convective accretion flow (see NIA for
an analysis using mixing length theory).  The reader satisfied with
the above argument may wish to proceed directly to \S4.

We consider a model in which the energy and angular momentum fluxes
have contributions from two sources: (1) hydrodynamic effects such as
bulk motion of the gas and convection and (2) non-hydrodynamic viscous
processes such as those due to magnetic fields.  We model the latter
with an $r\phi$-component of the stress tensor, which we denote $-T^v$,
where $T^v > 0$ because ``normal'' viscosity transports angular
momentum outwards.  In the usual Shakura-Sunyaev prescription, $T^v =
1.5 \nu \rho \vp/r$, where $\nu = \alpha c r$ is the kinematic
viscosity.

With this model, the radial fluxes can be written as \beq F_E =
 \la r^2 \rho v_r Be \ra  + r^2  \la \vp \ra T^v 
\label{eflux} \eeq \beq F_L =  \la \rho v_r r^3 \sin \theta \vp \ra
 + r^3 \sin \theta T^v, \label{lflux} \eeq \beq F_M = \la r^2 \rho v_r
\ra, \label{mflux} \eeq where $\la \ \ra $ denotes a time average over
many convective turnover times.  The first terms on the right hand
side of equations (\ref{eflux}) and (\ref{lflux}) are the hydrodynamic
contribution to the energy and angular momentum flux; these will be
explained in more detail below.  The second terms on the right hand
side of equations (\ref{eflux}) and (\ref{lflux}) are those due to the
viscous stress $T^v$.  Associated with the angular momentum flux due
to this stress is an energy flux $\propto \Omega T^v$ due to the rate
at which the stress does work on the gas ($\Omega = \vp/[r
\sin\theta]$ is the rotation rate of the gas).

To understand the meaning of the hydrodynamic contribution to the
angular momentum flux in equation (\ref{lflux}), we decompose $\vp$
into mean and fluctuating components via $\vp =  \la \vp \ra +
{\tilde \vp}$.  With this decomposition, \beq  \la \rho v_r r^3 
\sin \theta \vp \ra = r \sin\theta \la  \vp \ra F_M + r^3
\sin \theta T^c \label{lavg}, \eeq where $T^c =  \la \rho v_r
\tilde \vp \ra $ is the stress tensor associated with convection
(the Reynold's stress); $T^c$ is negative because convection is
assumed to transport angular momentum inwards.  Physically, the two
terms in equation (\ref{lavg}) correspond to the angular momentum flux
due to bulk motion and convection, respectively.

An analogous decomposition of the hydrodynamic contribution to the
energy flux in equation (\ref{eflux}) is less useful because the
energy flux contains a large number of quantities that are third order
in the fluctuating variables; a detailed analysis of these terms is
non-trivial and not particularly illuminating.  Instead, we simply
write \beq \la r^2 \rho v_r Be \ra = \la Be \ra F_M + r^2 \la \vp \ra
T^c + F_c
\label{eavg}, \eeq  which is formally just a definition of the convective 
energy flux $F_c$.  Physically, the first two terms on the right hand
side in equation (\ref{eavg}) correspond to the energy flux due to
bulk motion of the gas and the energy flux due to the rate at which
the ``convective stress,'' $T^c$, does work on the gas.  The remaining
term in equation (\ref{eavg}), $F_c$, is the energy flux due to
convection in the absence of bulk motion or angular momentum
transport.  This is, in fact, the energy flux usually considered in
convective media (e.g., mixing length theory).  In non-rotating stars,
e.g., convection does not lead to a significant transport of mass or
angular momentum, but it does lead to a significant flux of energy; in
our notation this is captured by the term $F_c$ in equation (\ref
{eavg}).  Although a detailed expression for $F_c$ in terms of the
turbulent quantities is complicated, it is also unnecessary for our
purposes.  All we need is its sign; $F_c > 0$ because convection
transports energy from high entropy (small radii) to low entropy
(large radii).

Using the Shakura-Sunyaev prescription for $T^v$, one finds that the
viscous contribution to the energy flux is $\propto r^{1/2 - n}$.
Similarly the viscous contribution to the angular momentum flux is
$\propto r^{2 - n}$.  In order to conserve energy, angular momentum,
and mass ($d{\bar F}/dr = 0$), the corresponding angle-integrated
fluxes must either be zero or independent of radius.  Therefore, we
can write our final conservation equations as \beq \bar F_E = 
\la Be \ra \bar F_M + \bar F_c + r^2  \la \vp \ra \bar T =
C_1 \delta_{n,1/2},
\label{efinal} \eeq \beq  \bar F_L = r  \la \vp \ra \bar F_M +
r^3 {\overline{T \sin\theta}} = C_2 \delta_{n,2},
\label{lfinal} \eeq \beq \bar F_M = \overline{\la r^2 \rho v_r \ra} 
= C_3 \delta_{n,3/2}, \label{mfinal} \eeq where the $C_i$ are
constants and $T = T^v + T^c$ is the total stress tensor.  In writing
equations (\ref{efinal})-(\ref{mfinal}) we have integrated the fluxes
on spherical shells and have used the fact that $ c^2 $, $ \vp^2 $,
and $ Be $ are independent of $\theta$ in the marginally stable state
and thus can be taken out of the angular integration.


Standard ADAF solutions correspond to $n = 3/2$.  As indicated by
equations (\ref{efinal})-(\ref{mfinal}), this is a solution for which
there is a finite flux of mass onto the central object, but no net
energy or angular momentum flux (Narayan \& Yi 1994; Blandford \&
Begelman 1999).  As argued above, no such solution is possible when
convection dominates the structure of the flow and drives it to
marginal stability.


Consider a putative $n = 3/2$ solution to equations
(\ref{efinal})-(\ref{mfinal}).  For such a solution $\bar F_M < 0$
since there is an inward flux of mass.  From equation (\ref{efinal}),
and using $ Be \approx 0$ near marginal stability, we see that $\bar
F_E \approx r^2 \vp \bar T + \bar F_c$.  Both terms in this expression
for $\bar F_E$ are positive and so it is not possible to have $\bar
F_E = 0$.\footnote{An inward flux of mass with $\bar F_L = 0$ requires
$\overline{T \sin\theta} > 0$.  In principle, one can have $\bar T <
0$ with $\overline{T \sin\theta} > 0$ since we do not know the
$\theta$ distribution of the convective ($T^c < 0$) and viscous ($T^v
> 0$) stresses (this would correspond to an outward angular momentum
flux and an inward energy flux due to ``viscosity + convection'').  It
is likely, however, that $\overline{T \sin\theta} > 0$ because $T > 0$
at all $\theta$, i.e., because viscosity wins out over convection
($|T^v| > |T^c|$) at all $\theta$.  In this case $\bar T > 0$
follows.}  There is consequently no solution with $n = 3/2$.  This
argument shows that an ADAF that is marginally stable to convection
has a net flux of energy to large radii. Mathematically, the
corresponding self-similar solution has $n = 1/2$.  In this solution,
$\bar F_M = 0$, i.e., there is no accretion; this is possible because
the inward transport of angular momentum by convection exactly
balances the outward transport by viscosity ($\overline{T \sin \theta}
= 0$).

\section{Discussion}

Our argument can be summarized as follows.  If convection in ADAFs
transports angular momentum inwards and if outward transport due to
other mechanisms is weak ($\alpha \ll 1$), the time averaged structure
of an ADAF must approach marginal stability to convection.  By
analyzing the structure of a marginally stable flow we find that there
must be a net outward flux of energy due to convection.  For a
self-similar flow there can either be a net flux of mass or a net flux
of energy, but not both (\S3).  Thus our marginally stable flow with a
finite energy flux has {\it zero} accretion rate in the self-similar
regime; the inward angular momentum transport due to convection
cancels the outward transport due to viscosity.  Physically, matter in
a given spherical shell circulates indefinitely in convective eddies
instead of accreting (in reality there will be a small net accretion
rate due to the violation of self-similarity near the surface of the
central object).  This marginally stable solution has a radial density
profile of $\rho \propto r^{-1/2}$, rather than the usual $\rho
\propto r^{-3/2}$ for spherical flows.

SPB and IA have carried out axisymmetric hydrodynamical simulations of
ADAFs; they find that the flow is convectively unstable for small
$\alpha$.  Our analytical calculation reproduces the time averaged
properties of these simulations rather well.  SPB's run K corresponds
to a Shakura-Sunyaev like viscosity with $\alpha = 10^{-3}$; in this
simulation, they find $\rho \propto r^{-1/2}$, as we derived in \S3.
SPB also consider non self-similar viscosity prescriptions for which
$\nu = \rho$ or $\nu = $ constant.  As discussed in \S3, for $\nu
\propto r^{\beta}$ a finite energy flux requires $n = \beta$ (where
$\rho \propto r^{-n}$).  This follows from the requirement that the
viscous energy flux $\propto r^2\nu \rho \vp ^2/r\propto \nu \rho$ be
independent of radius. Thus for both $\nu = \rho$ ($\beta = - n$) and
$\nu = $ constant ($\beta = 0$), a net energy flux requires $n = 0$,
i.e., $\rho$ independent of radius.  This is precisely what is found
in the simulations (see SPB's Fig. 7).

The interpretation of the numerical results as due to a marginally
stable flow is supported by the excellent quantitative agreement
between our results and the simulations.\footnote{SPB noted that the
angular structure of the contours of entropy, density, angular
momentum, and other dynamical variables found in the simulations
correspond to that expected near marginal stability (see the
discussion in Begelman \& Meier 1982).  Moreover, Igumenschev, Chen, \&
Abramowicz (1996) noted that convection in low $\alpha$ flows should
lead to a marginally stable state, which was consistent with the fact
that the surfaces of constant specific angular momentum and entropy
coincided in their simulations.} For example, SPB find rotational
velocities that are nearly constant on spherical shells (see their
Figs. 7 \& 12), as we derived in \S2 and the Appendix.\footnote{Near
the pole ($\theta \lsim 10^o$), $\vp$ decreases in the simulations.
This is because near the pole $\vp = $ constant corresponds to a
divergent rotation rate.  In reality, and in the simulations, this is
smoothed out by a $\theta-\phi$ component of the viscous stress tensor
not considered in this work.}  For $\gamma = 1.5$, our marginally
stable $n = 1/2$ solution has $c^2 = 0.22 v_K^2$, in good agreement
with IA's low $\alpha$ simulations, which find $c^2 \approx 0.2 v^2_K$
(see their Fig. 10).  In addition, our predicted angular density
profiles of $\rho \propto \sin^2 \theta$ (for $\gamma = 5/3$ and $n =
1/2$) and $\rho \propto \sin^3 \theta$ (for $\gamma = 5/3$ and $n =
0$) are in excellent agreement with the simulations (see SPB's Fig. 12
and Fig. 7, respectively), as is our prediction that the time averaged
Bernoulli constant should be small, i.e., $\ll v^2_K$ (see SPB's
Figs. 4 \& 9 and the discussion before their eq. [9]).

It is important to note that our analysis does not rely on the
suggestion of Blandford \& Begelman (1999) that ADAFs will drive
strong outflows.  In fact, our density profile of $\rho \propto
r^{-1/2}$ is derived assuming no outflows.

If relevant to real flows, the $\rho \propto r^{-1/2}$ density profile
of the convection-dominated accretion flow would have important
observational implications.  Most importantly, it would naturally
explain the low luminosities of many nearby supermassive black holes.
For a given density and temperature in the ISM of a host galaxy, the
accretion rate predicted by the convection-dominated accretion flow is
$\sim R_s/R_A$ times smaller than the Bondi accretion rate, where
$R_s$ is the Schwardschild radius of the black hole and $R_A \approx
GM/c_A^2$ is the accretion radius of the black hole ($c_A$ is the
sound speed of the gas at large radii).  This follows by evaluating
$\dot M = 4 \pi R^2 \rho v$ close to the black hole where the velocity
is of order the sound speed (the gas goes through a sonic point on its
way into the black hole).  Using $\rho \propto R^{-1/2}$ and $c
\propto R^{-1/2}$ then yields $\dot M \sim \dot M_B (R_s/R_A)$, where
the Bondi accretion rate is $\dot M_B \approx 4 \pi R_A^2 \rho_A c_A$
($\rho_A$ being the density of the gas in the vicinity of the
accretion radius).  Such a low accretion rate accounts naturally for
the low luminosity of supermassive black holes such as Sgr A* at the
center of our Galaxy.

In addition, the X-ray to radio flux ratio and the X-ray spectrum of
accreting supermassive black holes are sensitive to the variation of
density with radius, i.e., $n$ (Di Matteo et al. 2000; Quataert \&
Narayan 1999).  Smaller $n$, e.g., $n = 1/2$ instead of $n = 3/2$,
results in a larger X-ray to radio flux ratio and a much harder X-ray
spectrum (dominated by bremsstrahlung instead of Comptonization).
This appears to be consistent with several low-luminosity AGN in
nearby elliptical galaxies (Di Matteo et al. 2000).

There are, however, several concerns about the relevance of the
convection-dominated solution that can only be addressed by future
numerical simulations: (1) Is the statistical steady state we have
calculated unstable to non-axisymmetric perturbations?  This is
important because our conclusion that there must be a net energy flux
follows from the properties of the marginally stable state. (2) Does
radial convection transport angular momentum inwards in 3D flows?  If
not, one is likely to get a normal ADAF, perhaps modified by strong
outflows (Blandford \& Begelman 1999; see also IA's high $\alpha$
simulations).  Stone \& Balbus (1996) have argued that outward angular
momentum transport by convection requires coherent azimuthal pressure
gradients.  These are manifestly absent in the axisymmetric
simulations of SPB and IA.  Thus 2D convection is artificial in an
important way.  (3) How large is $\alpha$ due to magnetic fields?  If
$\alpha \gsim 0.1$, as is quite plausible, then convection may be a
minor perturbation since the gas flows into the black hole on a
timescale comparable to the convective turnover time.

\acknowledgements We are grateful to Ramesh Narayan for pointing out
that inward transport of angular momentum by convection was important
for understanding the numerical simulations of Stone et al., and for
several additional discussions.  We thank John Bahcall for comments on
the paper and the referee, Mitch Begelman, for several useful
comments.  AG was supported by the W. M. Keck Foundation and NSF
PHY-9513835. EQ acknowledges support provided by NASA through Chandra
Fellowship grant number PF9-10008 awarded by the Chandra X-ray Center,
which is operated by the Smithsonian Astrophysical Observatory for
NASA under contract NAS8-39073.

\newpage

\begin{appendix}

\section{Calculation of the Marginally Stable Solution}

In this Appendix we calculate the angular structure of a marginally
stable flow by analyzing equation (\ref{stability}) assuming radial
self-similarity.  Equation (\ref{stability}) can be written as the
inequality
\begin{equation}
Ax^2+By^2+Cxy \leq  0, \label{quad}
\end{equation}
where $x = dr$ and $y = r d\theta$.

The coefficients of the quadratic form are given by

\begin{equation}
A = g_{r} \partial_r s - {2 \gamma \vp \over R^2}  \sin\theta \partial_r (R \vp),
\end{equation}
\beq B = {g_{\theta} \over r} \partial_\theta s - { 2 \gamma \vp \over r R^2}
\cos{\theta} \partial_\theta (R \vp),\eeq
and
\beq C = g_{\theta} \partial_r s + {g_{r}
\over r} \partial_\theta s - {2 \gamma \vp \over R^2} \left(\cos\theta 
\partial_r 
(R \vp) + {\sin\theta \over r} \partial_\theta(R \vp)\right). \eeq

The marginally stable flow corresponds to
\begin{equation}
A\leq 0, ~~~~~~~~~~~ B\leq 0, ~~~~ {\rm and} ~~~~~~ C^2=4AB. \label{marg}
\end{equation}

The first two conditions follow because the flow must be stable to $x
= 0$ or $y = 0$ perturbations (purely tangential and purely radial,
respectively).  The condition $C^2 = 4AB$ follows because the left
hand side of equation (\ref{quad}) defines a quadratic function with
negative curvature.  If it has two real roots then it is positive
somewhere and hence there are perturbations which are unstable.
Marginal stability occurs when the two roots collapse to one ($C^2 =
4AB$) and stability happens when the two roots of the quadratic
function are imaginary ($C^2 < 4AB$).

In addition to marginal stability, the flow must satisfy radial and
$\theta$ momentum balance
\begin{equation}
{v^2_K \over r } = - \rho^{-1} \partial_r p + {\vp^2 \over r} \label{rad}
\end{equation}
and
\begin{equation}
\cot \theta \rho \vp ^2= \partial_\theta p, \label{ang}
\end{equation}
where $p = \rho c^2$ is the pressure and $c$ is the isothermal sound
speed.

With radial self-similarity, equations (\ref{rad}) and (\ref{ang})
become
\begin{equation}
1  = (1+n) c^2 + \vp^2
\end{equation}
and
\begin{equation}
\cot\theta \rho \vp^2 = c^2 \rho' + \rho (c^2)',
\end{equation}
where $'\equiv \partial_\theta $ and all dynamical variables (e.g., $\rho$,
$c^2$) are now dimensionless functions of $\theta$.

For a self-similar flow the coefficients of the quadratic form become
\begin{equation}
A= 1-n(\gamma - 1) + \vp^2 \left(n(\gamma - 1) - (\gamma + 1)\right),
\end{equation}
\begin{equation}
B= {\vp^2 \cot\theta \over 1 - \vp^2} \left((1 - \gamma) \cot \theta \vp^2 
(1 +n) - 2 \gamma \vp \vp'\right) - {2 \gamma \vp \over \sin^2\theta} 
\left(\vp \cos^2\theta + \vp' \cos\theta \sin\theta\right),
\end{equation}
and
\begin{equation}
C=2 \vp^2 \cot \theta \left( (\gamma - 1)n - (\gamma + 1)\right).
\end{equation}

The marginal stability requirement is that $C^2 = 4 A B$.  Defining a
new independent variable, $\zeta =-\ln \sin \theta $, this can be
written as an equation describing the marginally stable temperature
profile
\begin{equation}
{dc^2 \over d\zeta }={(c^2-c^2_1)(c^2_2-c^2)\over c^2_3-c^2},
\end{equation}
where
\begin{equation}
c^2_1={1\over {\gamma +1\over \gamma -1}-n},~~~~c^2_2={1\over
1+n},~~~~c^2_3={c^2_1+c^2_2\over 2}.
\end{equation}
Since the region extending from the equator to the pole corresponds to
$\zeta$ going from $0$ to $\infty$, the only non-singular solution
which extends from the equator to the pole has $c^2=c^2_1$.  Any other
solution becomes singular somewhere between the equator and the pole
or has a free surface at a finite $\theta$ (see below).

The above solution has $c^2$ and $\vp^2$ constant on spherical shells,
i.e., independent of $\theta$, with values

\begin{equation}
c^2={1\over {\gamma +1\over \gamma -1}-n}, ~~~~\vp^2=2{{1\over \gamma -1} 
-n\over
{\gamma +1\over \gamma -1}-n}. \label{solution2}
\end{equation}

From equation (\ref{ang}) the density is then a power law in
$\sin\theta$, with \beq \rho \propto (\sin \theta )^{2({1\over \gamma
-1}-n)}. \eeq

Equation (\ref{solution2}) gives the marginally stable state
corresponding to $C^2 = 4AB$. One can show that this state also
satisfies $A < 0$ and $B < 0$ and so all of the conditions in equation
(\ref{marg}) are satisfied.

The referee has pointed out, correctly, that there exists another
class of marginally stable solutions not considered in this paper.
These have a free surface at a finite $\theta$, where the density and
pressure vanish.  Outside of the free surface lies an empty funnel,
where a hydrodynamic or magnetohydrodynamic outflow may be driven.
The properties of these solutions are discussed in Blandford \&
Begelman (in preparation).

One final comment on convective stability is in order.  In evaluating
the stability of height integrated ADAFs to convection, Narayan \& Yi
(1994) and NIA used $N^2 + \kappa^2 > 0$ as their stability
requirement, where $N$ is the Brunt-Vaisala frequency and $\kappa$ is
the epicyclic frequency.  In our notation, this is equivalent to
taking $A < 0$ since, at the equator, $A = - \gamma(N^2 + \kappa^2)$.
In the full 2D flow, the criterion $A < 0$ is, however, a necessary,
but not a sufficient, criterion for convective stability (see the
discussion below eq. [\ref{quad}]).  Physically, $A < 0$ is the
stability requirement for radial perturbations, i.e., for $y \propto d
\theta = 0$ (see eq. [\ref{quad}]).  Near marginal stability, however,
the ``almost'' unstable modes have $y = - \sqrt{A/B} x$ and are thus
not radial perturbations.  True marginal stability therefore
corresponds to $C^2 = 4AB$ and not $A = 0$.  It turns out that our
marginally flow has a rotational velocity a factor of $\sqrt{2}$
larger than the flow defined by $A = 0$.\footnote{This additional
rotation is needed to stabilize all perturbations.}  NIA found that
their marginally stable flow (defined by $A = 0$) underpredicted the
rotational velocities found in IA's simulations by a factor of
$\approx \sqrt{2}$ (see the discussion in their \S7); this is readily
explained by the above analysis.\footnote{To be concrete, for $\gamma
= 5/3$, our $n = 1/2$ solution has $c^2 = 0.28 v^2_K$, while NIA's
solution gives $c^2 = 0.48 v^2_K$; the simulations find $c^2 \approx
0.25 v^2_K$, in excellent agreement with our result (see \S7 and
Fig. 4 of NIA).}

\end{appendix}

\end{document}